\documentclass[twocolumn,prb,aps,notitlepage]{revtex4}
\usepackage{tikz}
\usepackage{amssymb}
\usepackage{psfrag}
\usepackage{textcomp}
\usepackage{pdfsync}
\usepackage{amsmath}
\usepackage{amsfonts}
\usepackage{graphicx,bm,epstopdf}
\usepackage{subfigure}
\usepackage{comment}
\usepackage{verbatim}
\usepackage{color}
\usepackage{upgreek}
\usepackage{tikz}
\usepackage{natbib,hyperref}
\usepackage{ifpdf}
\usepackage{float}

\input{tcilatex}
\begin{document}

\title{Topological Hall effect in diffusive ferromagnetic thin films with
spin-flip scattering}
\author{Steven S.-L. Zhang$^1$}
\email{shulei.zhang@anl.gov}
\author{Olle Heinonen$^{1,2}$}
\email{heinonen@anl.gov}
\affiliation{$^1$Materials Science Division, Argonne National Laboratory, Lemont, Illinois
60439, USA\\$^2$Northwestern-Argonne Institute of Science and Technology, 2145 Sheridan Road, Evanston, Illinois 60208, USA}

\begin{abstract}
We study the topological Hall (TH) effect in a diffusive ferromagnetic metal
thin film by solving a Boltzmann transport equation in the presence of
spin-flip scattering. A generalized spin diffusion equation is derived,
which contains an additional source term associated with the gradient of the
emergent magnetic field that arises from skyrmions. Because of the source
term, spin accumulation may build up in the vicinity of the skyrmions. This
gives rise to a spin-polarized diffusion current that in general suppresses
the bulk TH current. Only when the spin diffusion length is much smaller
than the skyrmion size does the TH resistivity approach the value derived by
Bruno \textit{et al.} [Phys. Rev. Lett. \textbf{93}, 096806 (2004)]. We
derive a general expression of the TH resistivity that applies to thin-film
geometries with spin-flip scattering, and show that the corrections to the
TH resistivity become large when the size of room temperature skyrmions is
further reduced to tens of nanometers.
\end{abstract}

\maketitle




\section{Introduction}

In the last few decades, the notion of topological order has led to several
major breakthroughs in condensed matter physics. Along with the discovery of
momentum space topology that unveils a new phase of matter -- topological
insulators~\cite{Hasan&Kane10RMP_TI,Liang&scZhang11RMP_TI} -- the spotlight
has also been on magnetic skyrmions ~\cite%
{Nagaosa13Nat.Nano_Skyrmion,Fert17NatComm-Sk-review}, which are spin
textures possessing nontrivial real-space topology. Each magnetic skyrmion
carries an integer topological charge protected by a finite energy barrier~%
\cite{Miltat16PRB_Sk-TI-barrier}. The topological attributes endow skyrmions
with substantial robustness against boundaries and disorders during their
current-induced motion, which makes them promising candidates for electronic
applications~\cite%
{Fert17NatComm-Sk-review,yZhou16IEEE_skrym-device-revw,klWang17NanoLett_skyrm-device}%
.

After initially being identified in bulk chiral magnets~\cite%
{Muhlbauer09Sci_skyrmion,Yu10Nat_observ-SkX_helical-magnet}, intensive
efforts have been devoted to creating and manipulating nanoscale skyrmions
at room temperature in magnetic thin films and multilayer structures~\cite%
{Fert16Nat.Nano_add-DMI,Jiang15Sci-bubble-sk} that are more suitable for
practical applications. A central issue in any application is a scheme to
electrically detect magnetic skyrmions. In thin films and multilayers this
is usually based on the topological Hall (TH) effect~\cite{Bruno04PRL_THE}.
This effect arises from the Berry phase acquired by conduction electrons
when they traverse skyrmion textures and is therefore associated with
topological charges carried by the skyrmions. The TH effect was first
observed in bulk non-centrosymmetric magnetic materials such as MnSi~\cite%
{Tokura09PRL_Unusual-Hall_MnSi,09PfleidererPRL_THE_MnSi}, MnGe~\cite%
{Tokura11PRL_THE_MnGe} and \emph{etc.}, following the discovery of the
skyrmion lattice (SkX) phase in these materials. Later, TH measurements were
also carried out in chiral magnet thin films with more stable SkX phase~\cite%
{Chien12PRL_skyrm-FeGe,Tokura13PRL_THE_MnSi} than bulk systems. Recently, a
discretized TH effect induced by discontinuous motion, the creation and
annihilation of individual skyrmions, was observed in nanostructured FeGe
Hall-bar devices~\cite{Kanazawa15PRB_discret-THE}.

So far, all electrical measurements of magnetic skyrmions have been based on
the interpretation that the TH resistivity is proportional to the number of
skyrmions multiplied by the magnetic flux quantum $\psi _{0}\left( =\frac{h}{%
e}\right) $, \emph{i.e.}, $\rho _{yx}^{T}\propto N_{sk}\psi _{0}$, a result
originally derived by Bruno \emph{et al.}~\cite{Bruno04PRL_THE} for bulk
systems in the nondiffusive regime. We note that TH effect itself is \textit{%
not }topologically protected in the presence of scattering of the conduction
electrons. Indeed, Ndiaye and coworkers~\cite{Manchon17PRB_THE} recently
computed numerically the TH effect using the Landauer-B\"{u}ttiker
formulation and found it to be highly sensitive to spin-independent
impurities. Furthermore, previous studies of the TH effect have been focused
on the charge transport with conserved spin polarization, which only result
in a spin-polarization dependent prefactor in the TH resistivity. However,
this is not valid in the presence of spin accumulation and spin-flip
scattering: The former locally alters the spin polarization and the latter
mixes the two conduction channels for spin-up and spin-down electrons. These
observations raise an important question of what the effect of spin-flip
scattering is on the TH effect, and what the experimental consequences are
of spin accumulation in thin film geometries.

\begin{figure*}[tph]
\centering \hspace*{\fill}
\subfigure{
\includegraphics[trim={1cm 3cm 0.5cm 3cm},clip=true, width=0.35\linewidth]{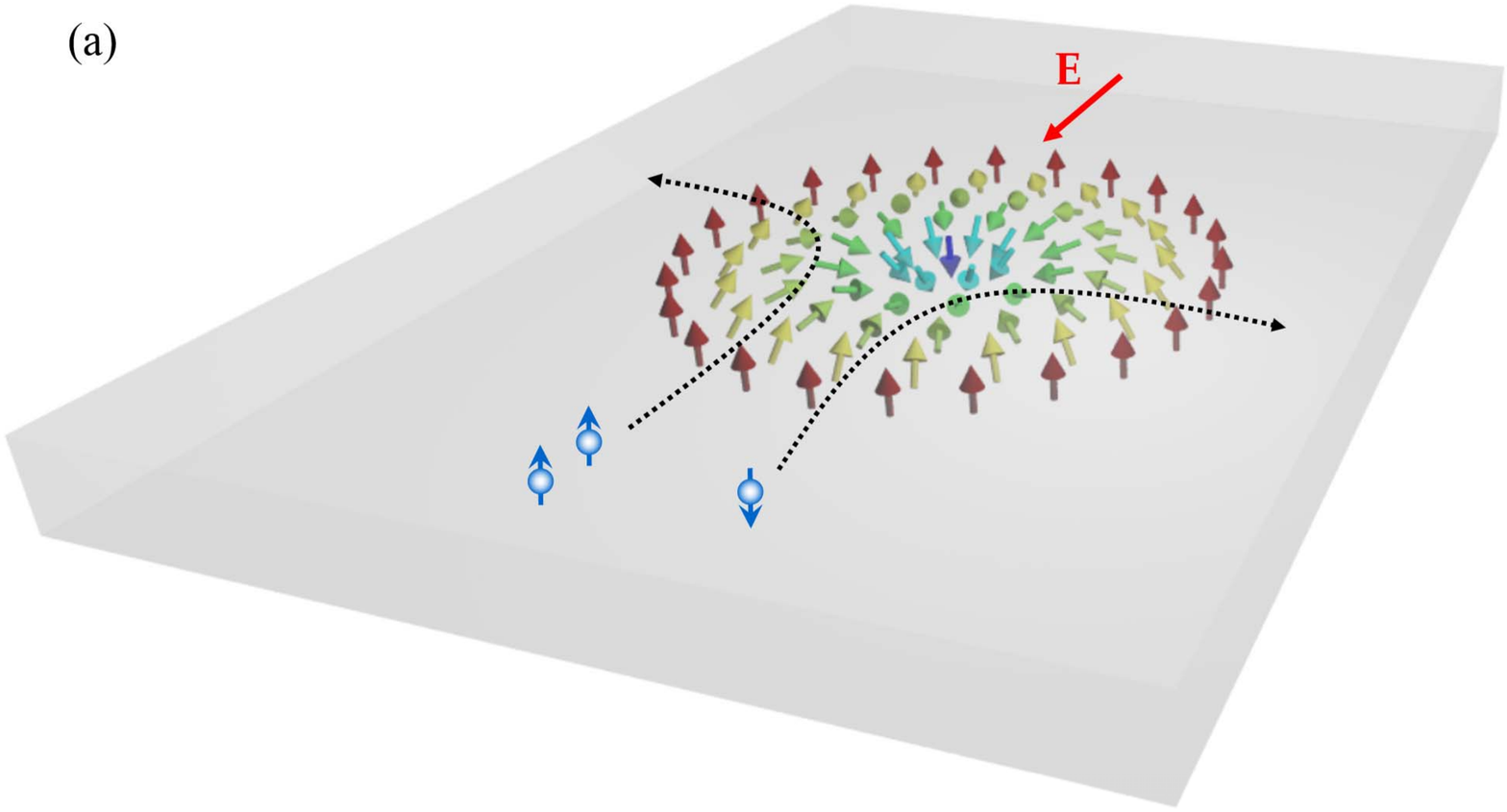}}
\quad
\subfigure{
\includegraphics[trim={1cm 3cm 0.5cm 3cm},clip=true, width=0.35\linewidth]{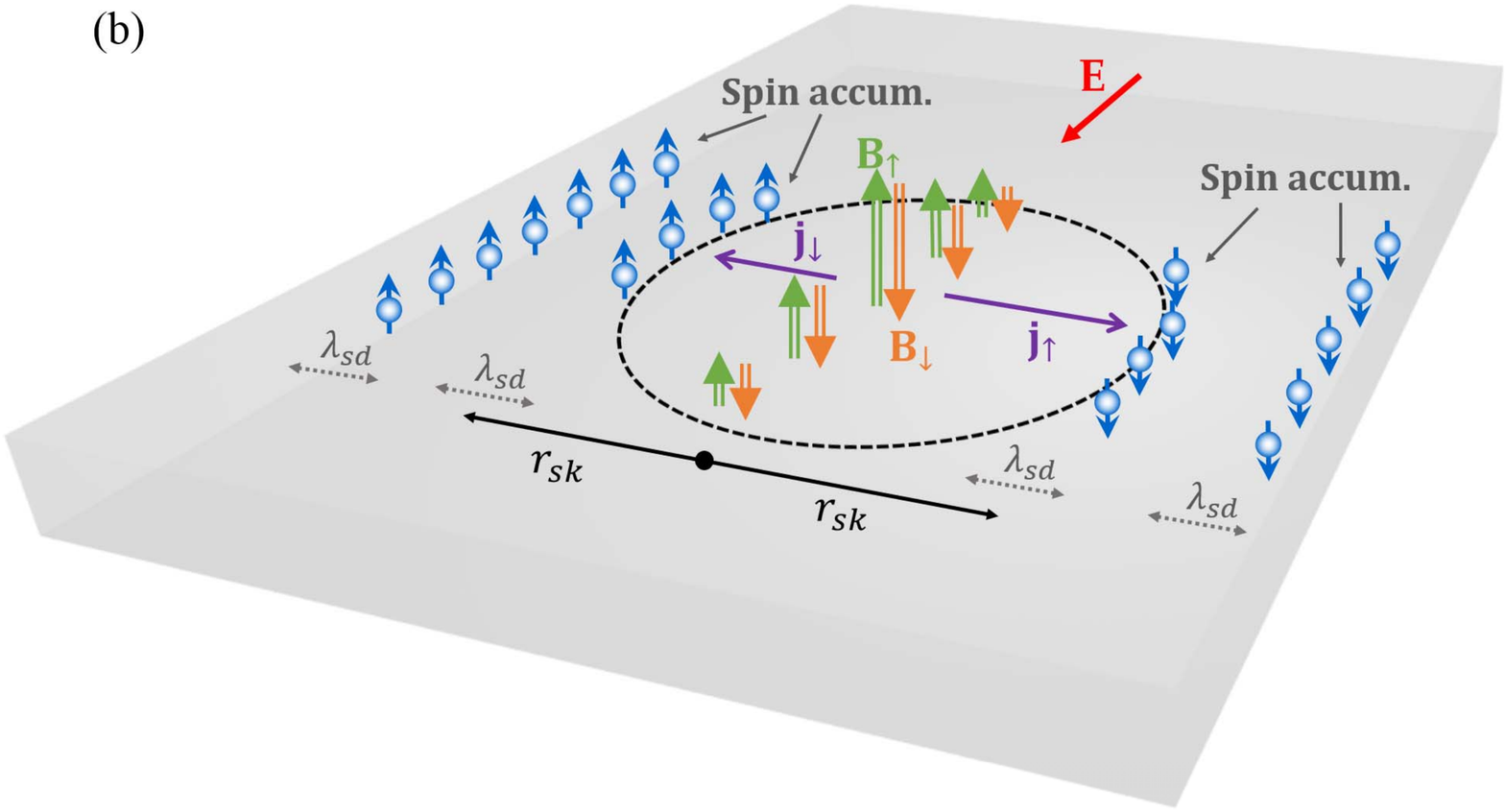}}
\hspace*{\fill}
\caption{Schematics of TH effect in a diffusive ferromagnetic thin film when an external electric field $\mathbf{E}$ is applied: (a) The trajectories of spin-up and spin-down electrons traversing a N\'{e}%
el-type skyrmion of radius $r_{sk}$ due to the TH effect and (b) spin accumulations built up at the lateral boundaries of the thin film as well as in the vicinity of the skyrmion within spin diffusion length $\lambda_{sd}$, owing to the nonuniform emergent magnetic field. Note that the emergent magnetic field acting on spin-up electrons, i.e., $\mathbf{B}_{\uparrow}$ (denoted by hollow green arrows), and that acting on spin-down electrons, i.e., $\mathbf{B}_{\downarrow}$ (denoted by hollow orange arrows), are of opposite directions.}
\label{fig:schematics}
\end{figure*}

Here, we investigate TH effect in a diffusive ferromagnetic metal (FM) thin
film with spin-flip scattering by treating spin and charge transport on an
equal footing. We show that skyrmions act as sources of spin accumulation,
which may build up not only near the lateral boundaries but also in the
vicinity of the skyrmions, as shown schematically in Fig.~\ref{fig:schematics}; consequently, the TH resistivity is in general
reduced and is no longer proportional to the number of skyrmions.

\section{The Semiclassical Transport Theory}

Our starting point is the steady-state Boltzmann transport equation
\begin{equation}
\mathbf{v\cdot }\frac{\partial f_{s}}{\partial \mathbf{r}}-e\left( \mathbf{E}%
+\mathbf{v\times B}_{s}\right) \cdot \frac{\partial f_{s}}{\hbar \partial
\mathbf{k}}=-\frac{f_{s}-\left\langle f_{s}\right\rangle }{\tau _{s}}-\frac{%
\left\langle f_{s}\right\rangle -\left\langle f_{-s}\right\rangle }{\tau
_{sf}}\,,  \label{Eq: Boltzmann}
\end{equation}%
where $f_{s}\left( \mathbf{r,k}\right) $ is the distribution function for
electrons with $s=1$ ($-1$) or $\uparrow $ ($\downarrow $) denoting spin-up
(spin-down) with respect to the local magnetization direction, $\left\langle
f_{s}\right\rangle \equiv \int d^{2}\Omega _{\mathbf{k}}$ $f_{s}\left(
\mathbf{r,k}\right) /\int d^{2}\Omega _{\mathbf{k}}$ with $\Omega _{\mathbf{k%
}}$ the solid angle in k-space, and $\tau _{s}$ and $\tau _{sf}$ are the
momentum and spin-flip relaxation times respectively; $\mathbf{E}$ is the
external electric field applied in the longitudinal direction of the FM thin
film, \textit{i.e.}, $\mathbf{E=}E_{x}\mathbf{\hat{x}}$, and $\mathbf{B}_{s}$
is the emergent magnetic field~\cite%
{Volovik87JPC,PAPANICOLAOU91_Nucl-Phys-B_Beff} given by
\begin{equation}
B_{s,i}=-s\frac{\hbar }{2e}\epsilon _{ijk}\mathbf{m}\cdot \left( \partial
_{j}\mathbf{m}\times \partial _{k}\mathbf{m}\right)  \label{Eq: Beff_s,i}
\end{equation}%
with $\mathbf{m}$\ the unit vector denoting the direction of the
magnetization, $\partial _{j}$ the short-hand denotation for the spatial
derivative of $\frac{\partial }{\partial r_{j}}$ and $\epsilon _{ijk}$ the
antisymmetric Levi-Civita tensor. We assume the film is sufficiently thin so
that the magnetization is uniform along the direction perpendicular to the
thin film (\textit{i.e.}, the $z$-direction), and focus on the transport in
the x-y plane. Note that, for this effectively two-dimensional system, only
the $z$-component of the emergent magnetic field is nonzero, i.e., $\mathbf{B%
}_{s}=sB\hat{z}$.

Note that by including the emergent magnetic field in Eq.~(\ref{Eq:
Boltzmann}) as a driving force of the conduction electrons, we have assumed
a strong exchange coupling $J_{ex}$ between the local magnetic
moments and conduction electron spins, i.e., $\tau _{s}J_{ex}/\hbar \gg 1$; in this case, the transverse spin component decays in a time scale
of $\tau _{ex}\sim \hbar /J_{ex}$ much shorter than the electron
momentum relaxation time and hence the spin of the conduction electrons can
be considered as adiabatically following the surrounding local moments.
Recently, THE in the weak exchange coupling regime has also been studied
theoretically~\cite%
{Denisov17Sci-Rep_THE-crossover,Kohno18JPSJ_THE-weak-coupling}. In this
work, we shall focus on the strong exchange coupling regime where the
emergent magnetic field (or the real-space Berry phase) picture is still
valid. Also note that, the emergent magnetic field was derived~\cite%
{Volovik87JPC,PAPANICOLAOU91_Nucl-Phys-B_Beff,sZhang09PRL,
Mecklenburg08PRB,sYang09PRL} by projecting the scalar and vector gauge
potentials -- emerging from a canonical transformation that rotates
quantization axis to align with the local magnetization -- onto the spin-up
and spin-down bands with respect to the local magnetization orientation; in
other words, the intrinsic interband spin-flip transition is neglected~\cite%
{sZhang10IEEE}. The spin-flip scattering considered in our transport
equation~(\ref{Eq: Boltzmann}) is of extrinsic origins such as
electron-magnon scattering and spin-orbit interactions with magnetic
impurities.

Next, we separate the distribution function into an equilibrium component $%
f_{0,s}\left( \mathbf{k}\right) $ and small nonequilibrium perturbations,
\begin{equation}
f_{s}\left( \mathbf{r,k}\right) =f_{0,s}\left( \mathbf{k}\right) -\frac{%
\partial f_{0,s}}{\partial \varepsilon _{ks}}\left[ -e\mu _{s}\left( \mathbf{%
r}\right) +g_{s}\left( \mathbf{r,k}\right) \right] \,,  \label{Eq: f-moments}
\end{equation}%
where $-e\mu _{s}\left( \mathbf{r}\right) $ and $g_{s}\left( \mathbf{r,k}%
\right) $ are the zeroth and first velocity moments respectively (the latter
satisfies $\int d^{2}\mathbf{k}$ $g_{s}\left( \mathbf{r,k}\right) =0$), and $%
\varepsilon _{ks}$ $=\frac{\hbar ^{2}k^{2}}{2m}-sJ_{ex}$ denotes the energy
of spin-$s$ electrons with $J_{ex}$ the exchange splitting of the conduction
band. Placing Eq.~(\ref{Eq: f-moments}) in the Eq.~(\ref{Eq: Boltzmann}) and
separating the odd and even velocity moments of the distribution function,
we find, up to $\mathcal{O}({B})$,%
\begin{equation}
g_{s}\simeq -e\tau _{s}\mathbf{v}\cdot \left( \mathbf{E}-\nabla _{\mathbf{r}%
}\mu _{s}-\frac{\tau _{s}e}{m}\mathbf{E\times B}_{s}\right) \,,
\label{Eq:g_s}
\end{equation}%
and a generalized diffusion equation for the spin accumulation, defined as $%
\delta \mu \equiv \frac{1}{2}\left( \mu _{\uparrow }-\mu _{\downarrow
}\right) $:
\begin{equation}
\nabla _{\mathbf{r}}^{2}\delta \mu -\frac{\delta \mu }{\lambda _{sd}^{2}}=%
\frac{\tau e}{m}\left( \mathbf{\hat{z}\times E}\right) \cdot \nabla _{%
\mathbf{r}}B\,.  \label{Eq:diff-eq-dmu}
\end{equation}%
Here, the spin averaged diffusion length $\lambda _{sd}$ is given by $%
\lambda _{sd}^{-2}\equiv \left( l_{\uparrow }^{-2}+l_{\downarrow
}^{-2}\right) /2$ with $l_{s}=\sqrt{\frac{1}{2}v_{F,s}^{2}\tau _{s}\tau _{sf}%
}$, $\tau =\frac{\tau _{\uparrow }+\tau _{\downarrow }}{2}$ is the spin
averaged momentum relaxation time, and we have used $\mathbf{B}_{+}=-\mathbf{%
B}_{-}=B\hat{z}$; detailed derivation of the generalized spin
diffusion equation is given in Appendix A. We note that the nonuniform
emergent magnetic field introduces a source term in the spin diffusion
equation, Eq.~(\ref{Eq:diff-eq-dmu}). In the absence of the source term,
spin accumulation can only occur at the boundaries through spin injection.
The source term gives rise to spin accumulation in the vicinity of the
skyrmions where the gradient of the emergent magnetic field is non-zero.
This imparts nonlocal features in the spin and charge transport, as we show
below.

The current density of the spin-$s$ conduction channel can be calculated by $%
\mathbf{j}_{s}=\frac{-e}{\left( 2\pi \right) ^{2}}\int d^{2}\mathbf{k}%
f_{s}\left( \mathbf{r,k}\right) \mathbf{v}$; using Eqs.~(\ref{Eq: f-moments}%
) and (\ref{Eq:g_s}), it can be expressed as
\begin{equation}
\mathbf{j}_{s}=\sigma _{s}\left( \mathbf{E}-\nabla _{\mathbf{r}}\mu
_{s}\right) -\sigma _{s}\left( \frac{\tau _{s}e}{m}\right) \mathbf{E}\,%
\mathbf{\times B}_{s},
\end{equation}%
where $\sigma _{s}=\frac{e^{2}\tau _{s}n_{e,s}}{m}$ is the longitudinal
conductivity with $n_{e,s}$ the density of spin-$s$ conduction electrons.
The charge and spin current densities by definition are given by the sum and
difference of the current densities in the two spin channels, i.e., $\mathbf{%
j}_{ch}=\mathbf{j}_{\uparrow }+\mathbf{j}_{\downarrow }$ and $\mathbf{j}%
_{sp}=\mathbf{j}_{\uparrow }-\mathbf{j}_{\downarrow }$.

The TH effect is associated with the transverse current densities that arise
from spatial variations of the chemical potentials along the $y$-direction.
Within linear response, the leading-order correction to the longitudinal
current due to gradients in the chemical potentials is of the second order
in the emergent magnetic field $\mathbf{B}_{s}$. Therefore up to $\mathcal{O}%
({B}$), we may focus only on the $y$-components of the charge and spin
current densities by integrating out the skyrmions along $x$-direction,
i.e.,
\begin{equation}
\bar{j}_{ch,y}=-\sigma \left( \frac{d\bar{\mu}}{dy}+p_{\sigma }\frac{d%
\overline{\delta \mu }}{dy}\right) +\sigma \left( p_{\tau }+p_{\sigma
}\right) E_{x}\left( \frac{\tau e\bar{B}}{m}\right)  \label{Eq:j_ch,y}
\end{equation}%
and
\begin{equation}
\bar{j}_{sp,y}\,=-\sigma \left( p_{\sigma }\frac{d\bar{\mu}}{dy}+\frac{d%
\overline{\delta \mu }}{dy}\right) +\sigma \left( 1+p_{\tau }p_{\sigma
}\right) E_{x}\left( \frac{\tau e\bar{B}}{m}\right) \,,  \label{Eq:j_sp,y}
\end{equation}%
where $\sigma =\sigma _{\uparrow }+\sigma _{\downarrow }$ and $\mu =\frac{%
\mu _{_{\uparrow }}+\mu _{_{\downarrow }}}{2}$ are the total longitudinal
conductivity and the spin averaged chemical potential, $p_{\sigma }\equiv
\frac{\sigma _{\uparrow }-\sigma _{\downarrow }}{\sigma _{\uparrow }+\sigma
_{\downarrow }}$ and $p_{\tau }\equiv \frac{\tau _{\uparrow }-\tau
_{\downarrow }}{\tau _{\uparrow }+\tau _{\downarrow }}$ are the spin
asymmetries of the conductivity and relaxation time respectively, and we
have defined $\bar{F}\left( y\right) \equiv \left( 2l\right)
^{-1}\int_{-l}^{l}dxF\left( \mathbf{r}\right) $ with $F\left( \mathbf{r}%
\right) $ denoting an arbitrary function and $2l$ the length of the film.
Similarly, we can cast the spin diffusion equation along the $y$ direction
as follows
\begin{equation}
\frac{d^{2}\overline{\delta \mu }}{dy^{2}}-\frac{\overline{\delta \mu }}{%
\lambda _{sd}^{2}}=E_{x}\left( \frac{\tau e}{m}\frac{d\bar{B}}{dy}\right) \,,
\end{equation}%
with the general solution given by
\begin{eqnarray}
\overline{\delta \mu }\left( y\right) &=&C_{-}\exp \left( -\frac{y}{\lambda
_{sd}}\right) +C_{+}\exp \left( \frac{y}{\lambda _{sd}}\right) -\lambda
_{sd}E_{x}  \notag \\
&&\times \int_{-w}^{+w}d\tilde{y}\left( \frac{\tau e}{2m}\frac{d\bar{B}}{d%
\tilde{y}}\right) \exp \left( -\frac{\left\vert y-\tilde{y}\right\vert }{%
\lambda _{sd}}\right) \,,  \label{Eq:dmu_sol}
\end{eqnarray}%
where $C_{\pm }$ are two constants of integration to be determined by the
boundary conditions, and $2w$ is the film width. Note that the spatially
varying emergent magnetic field induces a \textit{nonlocal} term in the spin
accumulation, with the range of nonlocality set by the spin diffusion
length. Similar nonlocal feature has also been found in the weak
exchange coupling regime~\cite{Kohno18JPSJ_THE-weak-coupling}. 

With open boundary conditions in the transverse direction, the transverse
current density in each spin-channel vanishes at the boundaries, i.e., $\bar{%
j}_{s,y}\left( \pm w\right) =0$, and it follows that $\bar{j}_{sp}\left( \pm
w\right) =0$. The transverse component of the charge current\ density, up to
$\mathcal{O}\left( B\right) $, must be zero everywhere, \textit{i.e.}, $\bar{%
j}_{ch,y}\left( y\right) =0$, due to the absence of bulk charge accumulation
in ferromagnetic metals with screening lengths of a few {\AA }ngstr\"oms. We
can thus express the transverse electric field $E_{y}$ in terms of the spin
accumulation and the emergent magnetic field as 
\begin{equation}
E_{y}=-\frac{d}{dy}\bar{\mu}=p_{\sigma }\frac{d}{dy}\overline{\delta \mu }%
-\left( p_{\tau }+p_{\sigma }\right) E_{x}\frac{e\tau \bar{B}}{m}\,.
\label{Eq:E_y}
\end{equation}%
Placing Eqs.~(\ref{Eq:j_ch,y}), (\ref{Eq:j_sp,y}) and (\ref{Eq:dmu_sol}) in
the boundary conditions, we can determine the spin accumulation $\overline{%
\delta \mu }\left( y\right) $ and hence $E_{y}$.

\section{Results and Discussions}

\begin{figure}[thp]
\centering
\subfigure{
\includegraphics[trim={0.8cm 0.5cm 3cm 2.2cm},clip=true, width=0.85\columnwidth]{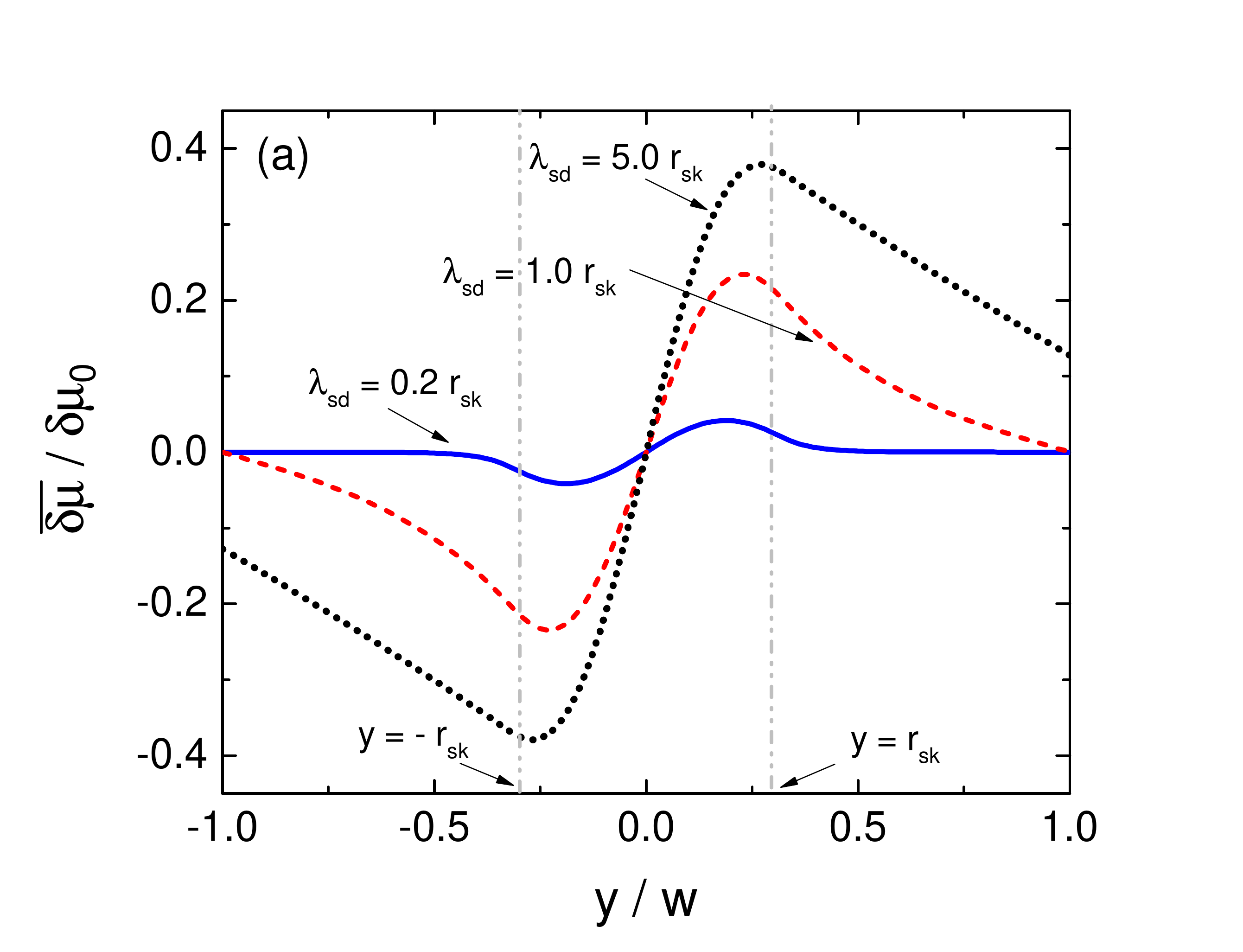}}
\hspace*{\fill}
\subfigure{
\includegraphics[trim={0.8cm 0.5cm 3cm 2.2cm},clip=true, width=0.85\columnwidth]{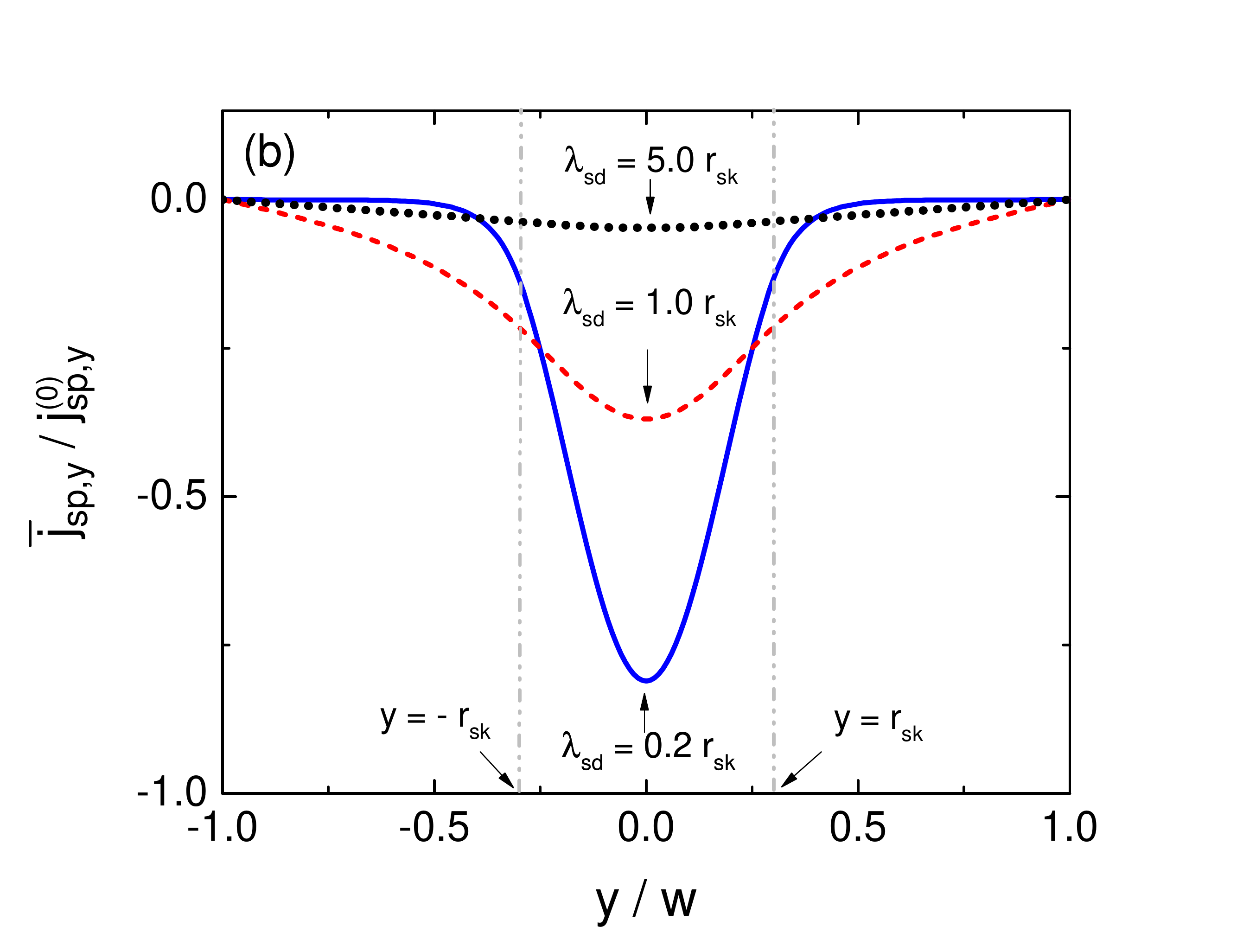}}
\hspace*{\fill}
\caption{Spatial profiles of (a) spin accumulation $\overline{\protect\delta
\protect\mu }$ and (b) transverse spin current density $\overline{j}_{sp,y}$
in the y-direction for several different spin diffusion lengths, where $%
\protect\delta \protect\mu _{0}\equiv E_{x}r_{sk}(\frac{e\protect\tau B_{0}}{%
m})$ and $j_{sp,y}^{(0)}\equiv (1-p_{\protect\sigma }^{2})\protect\sigma %
E_{x}(\frac{e\protect\tau B_{0}}{m})$, and $w=3r_{sk}$.}
\label{Fig:js-dmu}
\end{figure}

The magnitude of the TH effect can be characterized by the TH resistivity,
given by $\rho _{yx}^{T}=\frac{\bar{E}_{y}}{j_{x}}$ where $\bar{E}_{y}=\frac{%
1}{2w}\int_{-w}^{+w}dyE_{y}$. By placing Eq.~(\ref{Eq:dmu_sol}) in Eq.~(\ref%
{Eq:E_y}) and using the property that the total flux of the emergent
magnetic field associated with a skyrmion is equal to the magnetic flux
quantum $\psi _{0}$~\cite{Nagaosa13Nat.Nano_Skyrmion}, we obtain%
\begin{equation}
\rho _{yx}^{T}=\rho _{yx}^{T,\left( 0\right) }\left[ 1-\frac{p_{\sigma }}{%
p_{\tau }+p_{\sigma }}\int d^{2}\mathbf{r}\frac{B\left( \mathbf{r}\right)
\cosh \left( \frac{y}{\lambda _{sd}}\right) }{N_{sk}\psi _{0}\cosh \left(
\frac{w}{\lambda _{sd}}\right) }\right]   \label{Eq:rho_TH}
\end{equation}%
where $\rho _{yx}^{T,\left( 0\right) }=\left( p_{\tau }+p_{\sigma }\right)
n_{sk}R_{H}B_{0}$, $B_{0}=\frac{\psi _{0}}{\pi r_{sk}^{2}}$ is the averaged
emergent magnetic field per skyrmion with radius $r_{sk}$, $n_{sk}$ $=\frac{%
N_{sk}\pi r_{sk}^{2}}{4wl}$ is a dimensionless skyrmion density ($%
n_{sk}\rightarrow 1$ for a close packed SkX) with $N_{sk}$ the total number
of skyrmions contained in the thin film, $R_{H}=\frac{1}{en_{e}\left(
1+p_{\tau }p_{n}\right) }$ is the Hall coefficient with $n_{e}=n_{e,\uparrow
}+n_{e,\downarrow }$ and $p_{n}=\frac{n_{e,\uparrow }-n_{e,\downarrow }}{%
n_{e,\uparrow }+n_{e,\downarrow }}$ the total conduction electron density
and its spin polarization respectively. We notice that the TH
resistivity is proportional to the Hall coefficient and hence inversely
proportional to the carrier density; in other words, TH effect may become
more sizable in magnetic materials with lower carrier density. Also note
that, in deriving Eq.~(\ref{Eq:rho_TH}), we have assumed the effect of the
emergent magnetic fields from different skyrmions are additive, i.e., $%
B\left( \mathbf{r}\right) =\sum_{i}$ $B\left( \mathbf{r-r}_{i}\right) $ with
$\mathbf{r}_{i}$ denoting the central position of the $i$-th skyrmion~%
\footnote{%
This is true within linear response so long as the skyrmions are well
separated so that $2r_{sk}$ is smaller than the skyrmion-skyrmion separation.%
}.

Analytical results can be obtained in the following two limiting cases. When
$\lambda _{sd}\ll r_{sk}<w$, the TH resistivity coincides with the bulk
value, i.e., $\rho _{yx}^{T}\rightarrow \rho _{yx}^{T,\left( 0\right) }$~%
\cite{Bruno04PRL_THE} (provided $B\left( \mathbf{r}\right) $ has mirror
reflection symmetry about the $y=0$ plane), whereas in the opposite limit of
$\lambda _{sd}\gg w>r_{sk} $, we find the TH resistivity is reduced to $\rho
_{yx}^{T}\rightarrow \left( \frac{p_{\tau }}{p_{\tau }+p_{\sigma }}\right)
\rho _{yx}^{T,\left( 0\right) }$ . \ At a first glance, this observation may
seem a little counterintuitive as one would expect $\rho _{yx}^{T}$ to
approach its ideal bulk value in the weak diffusive limit (i.e., $\lambda
_{sd}\rightarrow \infty $). However, this is \textit{not} the case in the
presence of spin accumulation in thin FM films. To illustrate this point, it
is instructive to examine the spatial distributions of the spin current and
spin accumulation induced by a single skyrmion.

Let us consider a single skyrmion residing in the center of the thin film.
The magnetization unit vector defining a skyrmion may be expressed as $%
\mathbf{m}\left[ \Theta \left( r\right) ,\Phi \left( \phi \right) \right]
\mathbf{=}\left( \sin \Theta \cos \Phi ,\sin \Theta \sin \Phi ,\cos \Theta
\right) $\thinspace , where $\Theta $ and $\Phi $ are the polar and
azimuthal angles of the magnetization at a given position $\mathbf{r=}%
r\left( \cos \phi ,\sin \phi \right) $ relative to the center of the
skyrmion. In FM thin films with perpendicular anisotropy in contact with a
spin-orbit scatterer such as Ta or Pt, interfacial Dzyaloshinskii-Moriya
interactions favor N\'{e}el-type skyrmions~\cite%
{Sampaio13NatNano_Neel-Skyrm,Boulle16NatNano_sk-thinFilm,Jiang15Sci-bubble-sk,Woo16NatMater_sk-thinFilm}%
, for which the azimuthal angle can be written as $\Phi \left( \phi \right)
=\nu \phi +\gamma _{c}$ with vorticity $\nu =1$ and chirality $\gamma
_{c}=\pm \pi $ and the polar angle of the magnetization -- varying linearly
with its distance from the center of the skyrmion -- may be described as $%
\Theta \left( r\right) =\pi \left( 1-\frac{r}{r_{sk}}\right) H\left(
r_{sk}-r\right) \,$with $H\left( x\right) $ is the unit step function.

In Fig.~\ref{Fig:js-dmu}, we show the spatial distributions of the spin
accumulation $\overline{\delta \mu} $ and transverse spin current density $%
\overline{j}_{sp,y}$ along the $y$-direction of the thin film (averaged over
the $x$-coordinate across the skyrmion). When $\lambda _{sd}\ll r_{sk}$,
because of the short spin diffusion length both $\overline{\delta \mu }$ and
$\bar{j}_{sp,y}$ are spatially localized to within the skyrmion texture, and
hence the boundaries have no effect on the spin transport. However, when $%
\lambda _{sd}\ $becomes comparable to or larger than $r_{sk}$, both $%
\overline{\delta \mu }$ and $\bar{j}_{sp,y}\,$ spread out over the spin
diffusion length. It follows that the diffusive spin current generated by
the gradient of the spin accumulation will be partially converted to a
charge current due to the spin asymmetry of conductivities for spin-up and
spin-down electrons (\textit{i.e.}, $p_{\sigma }\neq 0$), which consequently
suppresses the bulk contribution of the TH current, as indicated by Eq.~(\ref%
{Eq:E_y}). Similar nonlocal spin and charge transport driven by the emergent
electric field was studied earlier \cite{sZhang10PRB} in the context of the
spin electromotive force~\cite{Barnes07PRL,sYang09PRL} and enhanced damping
induced by magnetization dynamics~\cite{sZhang09PRL,sZhang10IEEE}.

\begin{figure}[H]
\includegraphics[trim={0.8cm 0.5cm 3cm 2.2cm},clip=true,
width=0.85\columnwidth]{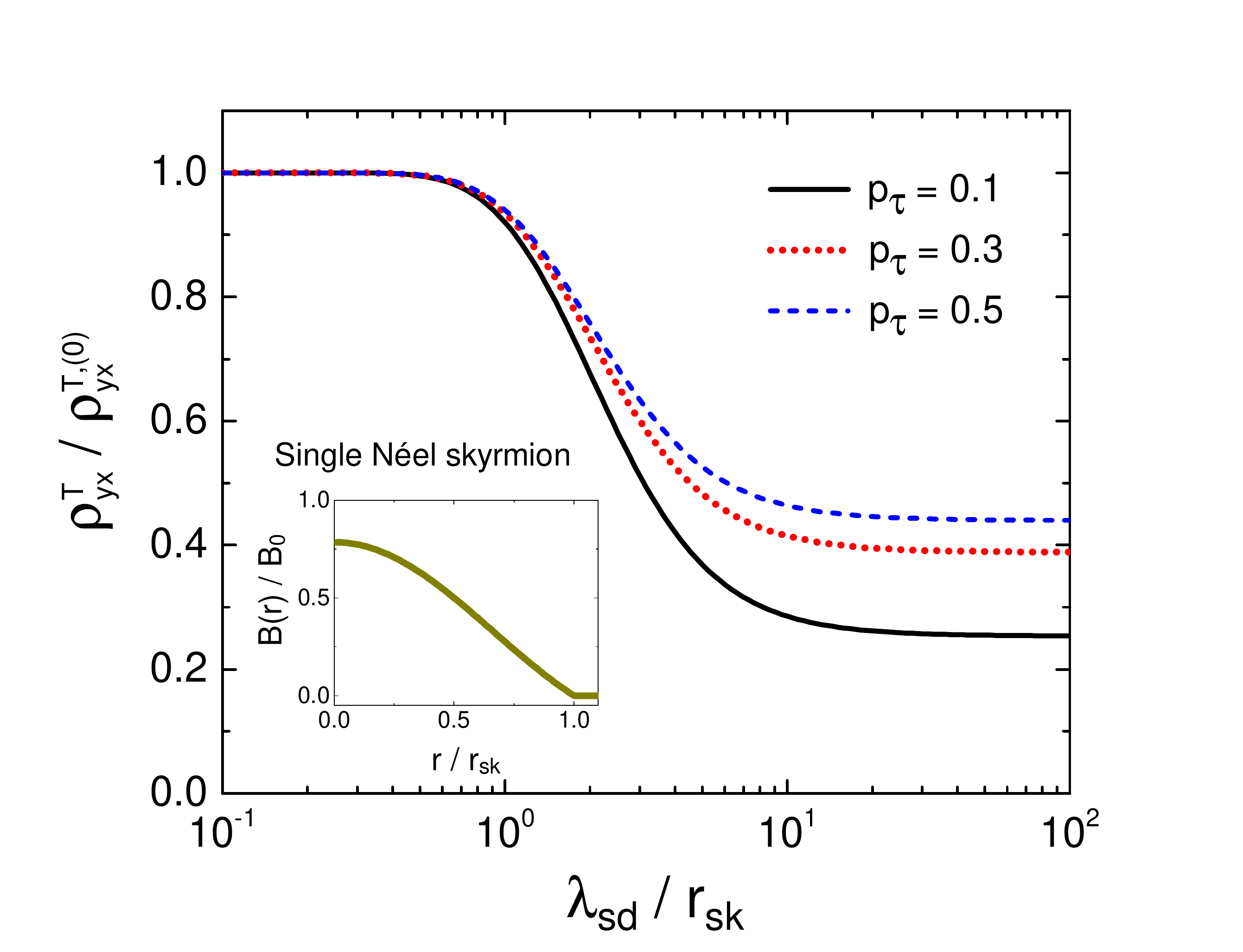}
\hspace*{\fill}
\caption{TH resistivity $\protect\rho _{yx}^T$ generated by a single N\'{e}%
el skyrmion as a function of spin diffusion length $\protect\lambda _{sd}$
in a thin film of half width $w=3r_{sk}$ for several different $p_{\protect%
\tau }$, where $\protect\rho _{yx}^{T,(0)}$ is the ideal bulk TH resistivity
independent of $\protect\lambda _{sd}$ ; the insets show the corresponding
spatial profile of the emergent magnetic field as a function of the distance
from the skyrmion center. Other parameters used: $J_{ex}/\protect\varepsilon%
_F=0.2$~\protect\cite{ptau}.}
\label{Fig:rho-1Neel-Skrm}
\end{figure}

Using Eq.~(\ref{Eq:rho_TH}), we can also numerically calculate the TH
resistivity induced by a N\'{e}el skyrmion for any arbitrary magnitude of
the spin diffusion length, as shown in Fig.~\ref{Fig:rho-1Neel-Skrm}. We
find that the TH resistivity decreases with increasing spin diffusion
length, reaching its maximum and minimum values in the two limits of $%
\lambda _{sd}/r_{sk}\rightarrow 0$ and $\lambda _{sd}/r_{sk}\rightarrow
\infty $ respectively; this is because longer spin diffusion length leads to
larger gradient of spin accumulation and hence greater backflow spin
(polarized) diffusion current that effectively suppresses the source current
generated by the skyrmion, as indicated by Fig.~\ref{Fig:js-dmu}. For given
ratios of $\lambda _{sd}/r_{sk}$ and $J_{ex}/\varepsilon _{F}$, $\rho
_{yx}^{T}$ decreases with the spin asymmetry of the electron momentum
relaxation $p_{\tau }$. We also note that, with fixed topological charge $%
\nu $ and the materials parameters $\lambda _{sd}/r_{sk}$ and $p_{\tau }$,
the TH resistivity is \textit{independent} of the detailed spin
configuration of the skyrmion; the analyses of the TH effect for a
Bloch-type skyrmion and a chiral skyrmion bubble with the same topological
charge can be found in Appendix B.

\begin{figure}[H]
\includegraphics[trim={1.3cm 0.4cm 1.2cm 2.3cm},clip=true,
width=0.85\columnwidth]{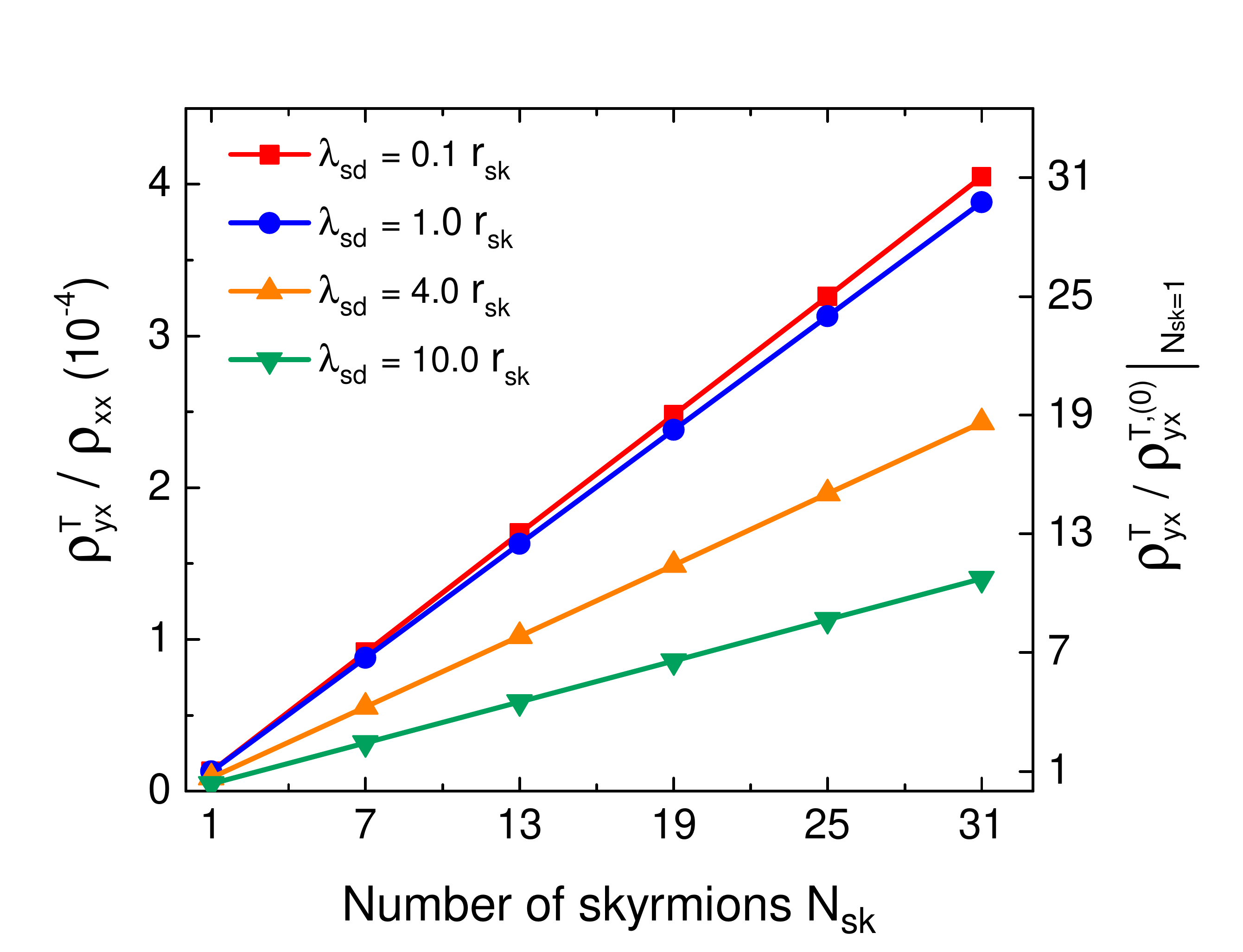}
\caption{The ratios of $\protect\rho_{yx}^T/\protect\rho_{xx}$ and $\protect%
\rho_{yx}^T/\left.\protect\rho_{yx}^{T,(0)}\right \vert_{N_{sk}=1} $ as a
function of skyrmion number for several spin diffusion lengths $\protect%
\lambda _{sd}$, where $\protect\rho_{xx}$ is the longitudinal resistivity
and $\left.\protect\rho_{yx}^{T,(0)} \right \vert_{N_{sk}=1} $ is the ideal
TH resistivity for a single skyrmion. Parameters used: $p_{\protect\tau %
}=0.1 $, $J_{ex}/\protect\epsilon _{F}=0.2$, $\protect\tau =10^{-13}$~s, $%
m=10^{-30}$~kg, $w=6r_{sk}$, and inter-skyrmion distance $d_{sk-sk}=4~r_{sk}$%
.}
\label{Fig:rho-Neel-SkX}
\end{figure}

Having understood the TH effect in the diffusive regime for a single
skyrmion, we now proceed to estimate the TH resistivity induced by a
hexagonal SkX. Figure~\ref{Fig:rho-Neel-SkX} shows the TH resistivity $\rho
_{yx}^{T}$ as a function of the number of skyrmions $N_{sk}$ in FM thin
films with different spin diffusion lengths. We find that $\rho _{yx}^{T}$
increases linearly with the number of skyrmions contained in the thin film,
as the emergent magnetic field of different skyrmions are additive. However,
only in the limit of short spin diffusion length (\textit{i.e.}, $\lambda
_{sd}/r_{sk}\rightarrow 0$), is the TH resistivity proportional to the
number of skyrmions, in agreement with the ideal bulk TH effect~\cite%
{Bruno04PRL_THE}. For a close-packed skyrmion lattice (\textit{i.e.}, $%
n_{sk}\rightarrow 1$) with $r_{sk}=100$~nm\ and a spin diffusion length of $%
\lambda _{sd}=10$~nm~\cite%
{jBass07JPhysCondMatt_spin-flip,Pana2017NatMater_skyrm-IrFeCoPt}, the
estimated TH resistivity is about $\rho _{yx}^{T}\sim 0.016~\mu \Omega \cdot
cm$, where we have used $R_{H}=0.05~\mu \Omega \cdot$ cm and $p_{\sigma
}=2p_{\tau }=0.4$. This is in order-of-magnitude agreement with recent
measurements of the TH resistivity in transition metal thin films and
multilayers~\cite{Chien12PRL_skyrm-FeGe,Pana2017NatMater_skyrm-IrFeCoPt}.
When the skyrmion size is further reduced, the magnitude of the TH
resistivity depends on the trade-off between the increased emergent magnetic
field $B_{0}$\ and the reduction due to spin accumulation. 
For example, with the same material parameters but a smaller skyrmion radius
$r_{sk}=0.5\lambda _{sd}=5$~nm\textbf{, }$\rho _{yx}^{T}$ will be reduced by
40\%; in the extreme case of $p_{\tau }/p_{\sigma }\rightarrow 0$\ and $%
r_{sk}/\lambda _{sd}\rightarrow 0$, $\rho _{yx}^{T}$\ might even be
completely suppressed.

As a final point, we discuss material considerations in observations of the
TH effect in FM thin films and multilayers. The spin diffusion length of
typical transition metal ferromagnets, typically of the order of $10$~nm~%
\cite{jBass07JPhysCondMatt_spin-flip}, is much smaller than the room
temperature skyrmions (with size in the range of $0.1\sim 1~\mu m$) observed
in magnetic thin films~\cite%
{Jiang15Sci-bubble-sk,Fert16Nat.Nano_add-DMI,Phatak16NanoLett_skyrm-multiferro, sZhang17APL_skyrm-NMG}%
. In this case, the spin and charge transport remain spatially localized to
within the skyrmion spin texture, and the measured TH resistivity will
therefore still agree with the ideal bulk TH effect~\cite{Bruno04PRL_THE},
as shown by Figs.(\ref{Fig:js-dmu}) and~(\ref{Fig:rho-1Neel-Skrm}). However,
increasing efforts have been directed at searching for nanoscale skyrmions
at room temperature, which is desirable for device applications~\cite%
{Fert16Nat.Nano_add-DMI}. According to our theory, when the skyrmion size
approaches the spin diffusion length of the FM layer, a reduction of the TH
resistivity should be expected, as indicated by Eq.~(\ref{Eq:rho_TH}). In
order to minimize such reduction, FMs with small $p_{\sigma }$ but sizable $%
p_{\tau }$ would be advantageous. For transition metals, $p_{\sigma }$ is in
fact dominated by scattering of the $s$-electrons from the $3d$-shell, which
can be tuned in their alloys. For example, Fert and Campbell~\cite%
{Fert76JPhysF_alloys} have demonstrated that in Ni-based binary alloys, $%
p_{\sigma }$ may be varied significantly and even tuned to change signs.

\section{Summary and Conclusion}

In this work, we exploited a semiclassical spin-dependent Boltzmann
equation to study the effect of spin-flip scattering and spin accumulation
on the TH effect in ferromagnetic thin films. We found that the nonuniform
emergent magnetic field serves as a source term for the spin diffusion and
imparts nonlocal features to the spin and charge transport. A generalized
spin diffusion equation was derived, whose solution shows that spin
accumulation may build up not only at the lateral boundaries of a current
carrying thin film but also in the vicinity of the magnetic skyrmions; such
spin accumulation gives rise to a spin-polarized diffusion current that
flows against the bulk TH current and hence attenuates the TH signal. A
general expression for the TH resistivity was obtained which applies to
diffusive ferromagnetic thin films especially when the size of the magnetic
skyrmions is comparable or smaller than the spin diffusion length. 

\bigskip

\bigskip

\section*{ACKNOWLEDGMENT}

One of the authors (S.S.-L.Z.) would like to thank Albert Fert, Alireza
Qaiumzadeh and Kyoung-Whan Kim for helpful discussions. This work was supported by the
Department of Energy, Office of Science, Basic Energy Sciences, Materials
Sciences and Engineering Division.

\appendix

\section{Derivation of the generalized spin diffusion equation}

Let us begin with the following Boltzmann equation
\begin{equation}
\mathbf{v\cdot }\frac{\partial f_{s}}{\partial \mathbf{r}}-e\left( \mathbf{E}%
+\mathbf{v\times B}_{s}\right) \cdot \frac{\partial f_{s}}{\hbar \partial
\mathbf{k}}=-\frac{f_{s}-\left\langle f_{s}\right\rangle }{\tau _{s}}-\frac{%
\left\langle f_{s}\right\rangle -\left\langle f_{-s}\right\rangle }{\tau
_{sf}}\,,  \tag{A1}  \label{Eq-APP: Boltzmann}
\end{equation}%
where $f_{s}\left( \mathbf{r,k}\right) $ is the distribution function for
electrons with spin character $s$, $\left\langle f_{s}\right\rangle \equiv
\int d^{2}\Omega _{\mathbf{k}}$ $f_{s}\left( \mathbf{r,k}\right) /\int
d^{2}\Omega _{\mathbf{k}}$ with $\Omega _{\mathbf{k}}$ the solid angle in
k-space, and $\tau _{s}$ and $\tau _{sf}$ are the momentum and the spin-flip
relaxation times respectively; $\mathbf{E}$ is the external electric field
is applied along the longitudinal direction of the FM thin film, \textit{i.e.%
}, $\mathbf{E=}E_{x}\mathbf{\hat{x}}$, and $\mathbf{B}_{s}$ is the emergent
magnetic field. Separating the distribution function into an equilibrium
component $f_{0,s}\left( \mathbf{k}\right) $ and small nonequilibrium
perturbations as follows
\begin{equation}
f_{s}\left( \mathbf{r,k}\right) =f_{0,s}\left( \mathbf{k}\right) -\frac{%
\partial f_{0,s}}{\partial \varepsilon _{ks}}\left[ -e\mu _{s}\left( \mathbf{%
r}\right) +g_{s}\left( \mathbf{r,k}\right) \right] \,,  \tag{A2}
\label{Eq-APP: f-moments}
\end{equation}%
where $e\mu _{s}\left( \mathbf{r}\right) $ and $g_{s}\left( \mathbf{r,k}%
\right) $ are the zeroth and first velocity moments respectively (the latter
satisfies $\int d^{2}\mathbf{k}$ $g_{s}\left( \mathbf{r,k}\right) =0$), and $%
\varepsilon _{ks}$ $=\frac{\hbar ^{2}k^{2}}{2m}-sJ_{ex}$ denotes the energy
of spin\thinspace -$s$ electrons with $J_{ex}$ the exchange splitting of the
conduction band. Placing Eq.~(\ref{Eq: f-moments}) in the Eq.~(\ref{Eq-APP:
Boltzmann}) and separating the odd and even velocity moments of the
distribution function, we find, up to $O\left( \mathbf{B}_{s}\right) $,

\begin{equation}
-e\left( \mathbf{E}-\nabla _{\mathbf{r}}\mu _{s}\right) \cdot \mathbf{v}%
+e\left( \mathbf{v\times B}_{s}\right) \cdot \frac{\partial g_{s}}{\partial
\mathbf{k}}=\frac{g_{s}}{\tau _{s}}  \tag{A3}  \label{Eq-APP:ani-terms}
\end{equation}%
and
\begin{equation}
\bigskip \mathbf{v\cdot }\frac{\partial }{\partial \mathbf{r}}g_{s}\left(
\mathbf{r,k}\right) =\frac{e\left( \mu _{s}-\mu _{-s}\right) }{\tau _{sf}}\,,
\tag{A4}  \label{Eq-APP:g-mu}
\end{equation}%
where we have assumed a spherical Fermi surface and have neglected
higher order terms of the order of $O\left( \frac{J_{ex}\tau _{s}}{%
\varepsilon _{F}\tau _{sf}}\right) $ [noting that $\frac{J_{ex}}{%
\varepsilon _{F}}\sim 0.1$ for typical transition metal
ferromagnets and $\frac{\tau _{s}}{\tau _{sf}}\ll 1$ for the
diffusive regime]. By placing the ansatz $g_{s}=\mathbf{A}_{s}\mathbf{\cdot
v}$ (where $\mathbf{A}_{s}$ is an arbitrary vector independent of $\mathbf{v}
$) in Eq.~(\ref{Eq-APP:ani-terms}) and solving the resulting vector equation
for $\mathbf{A}_{s}$, up to $O\left( \mathbf{B}_{s}\right) $, we find
\begin{equation}
g_{s}\simeq -e\tau _{s}\mathbf{v}\cdot \left( \mathbf{E}-\nabla _{\mathbf{r}%
}\mu _{s}-\frac{\tau _{s}e}{m}\mathbf{E\times B}_{s}\right) \,,  \tag{A5}
\label{Eq-APP: sol-g_s}
\end{equation}%
Note that since spatial variation in chemical potential is induced by $%
\mathbf{B}_{s}$, so we have discarded the higher order term\ $\nabla _{%
\mathbf{r}}\mu _{s}\mathbf{\times B}_{s}$ as well. Plugging Eq.~(\ref%
{Eq-APP: sol-g_s}) back in Eq.~(\ref{Eq-APP:g-mu}) and carrying out the
angular integration in the momentum space on both sides of the resulting
equation, we arrive at%
\begin{equation}
\nabla _{\mathbf{r}}^{2}\mu _{s}-\frac{\mu _{s}-\mu _{-s}}{l_{s}^{2}}=-\frac{%
\tau _{s}e}{m}\mathbf{E\cdot }\left( \nabla _{\mathbf{r}}\times \mathbf{B}%
_{s}\right) \,,  \tag{A6}  \label{Eq-APP:diff-eqn-mu_s}
\end{equation}%
where $l_{s}=\sqrt{\frac{1}{2}v_{F,s}^{2}\tau _{s}\tau _{sf}}$. Subtracting
Eq.~(\ref{Eq-APP:diff-eqn-mu_s}) for $s=$ $\uparrow $ from that for $s=$ $%
\downarrow $, we arrive at the generalized spin diffusion equation%
\begin{equation}
\nabla _{\mathbf{r}}^{2}\delta \mu -\frac{\delta \mu }{\lambda _{sd}^{2}}=%
\frac{\tau e}{m}\left( \mathbf{\hat{z}\times E}\right) \cdot \nabla _{%
\mathbf{r}}B\,,  \tag{A7}  \label{Eq-APP:gen-diffusion}
\end{equation}%
where the spin averaged diffusion length $\lambda _{sd}$ is given by $%
\lambda _{sd}^{-2}\equiv \left( l_{\uparrow }^{-2}+l_{\downarrow
}^{-2}\right) /2$, $\tau =\frac{\tau _{\uparrow }+\tau _{\downarrow }}{2}$
is the spin averaged momentum relaxation time and we have used the relation $%
\mathbf{B}_{+}=-\mathbf{B}_{-}=B\hat{z}$. Note that the generalized spin
diffusion equation (\ref{Eq-APP:gen-diffusion}) involves an additional
source term associated with the spatial gradient of the emergent magnetic
field. \bigskip

\section{Topological Hall effect induced by skyrmions with different spin
configurations}

\subsection{Bloch-type skyrmion}

A standard Bloch-type skyrmion has the same polar angle profile as a
standard N\'{e}el-type skyrmion but with a different chirality of $\gamma
_{c}=\pm \frac{\pi }{2}$. However, the emergent magnetic field does not rely
on $\gamma _{c}$, as can be explicitly seen from the expression of the
emergent magnetic field in polar coordinate system ~\cite%
{Nagaosa13Nat.Nano_Skyrmion} , \textit{i.e.},
\begin{equation}
B\left( r\right) =-\frac{\hbar }{2e}\sin \Theta \frac{1}{r}\frac{d\Theta }{dr%
}\frac{d\Phi }{d\phi }  \tag{B1}
\end{equation}
It follows that the topological Hall resistivity of Bloch-type skyrmions (or
skyrmion lattices) should be the same as that of N\'{e}el-type skyrmions (or
skyrmion lattices) which was examined in the main text.

\subsection{Chiral skyrmion bubble}

Skyrmion bubbles (or chiral bubbles) were also found in magnetic thins and
multilayers in recent experiments~\cite%
{Jiang15Sci-bubble-sk,heinonen16PRB_sk-SH}. At variance with a prototypical N%
\'{e}el skyrmion, the domain of reversed magnetization in a skyrmion bubble
is more extended and surrounded by a narrow N\'{e}el domain wall with fixed
chirality. The magnetization polar angle profile may be expressed as $\Theta
\left( r\right) =\frac{\pi }{2}\left\{ 1-\tanh \left[ f\left( r\right) %
\right] \,\right\} $~\cite{DeBonte73JAP_skym-bubbl-profile} with $f\left(
r\right) =\ln \left( \frac{2r}{r_{sk}}\right) +\frac{2r-r_{sk}}{2a_{DW}}$
and $a_{DW}$ the N\'{e}el wall width. The emergent magnetic field for a
skyrmion bubble can be written as%
\begin{equation}
B_{z}=\frac{\pi \hbar }{4e}\frac{\sin \Theta f^{\prime }\left( r\right) }{%
\left( \cosh \left[ f\left( r\right) \right] \right) ^{2}r}  \tag{B2}
\end{equation}

\begin{figure}[b]
\centering
\subfigure{
\includegraphics[trim={0.8cm 0.5cm 3cm 2.2cm},clip=true, width=0.85\columnwidth]{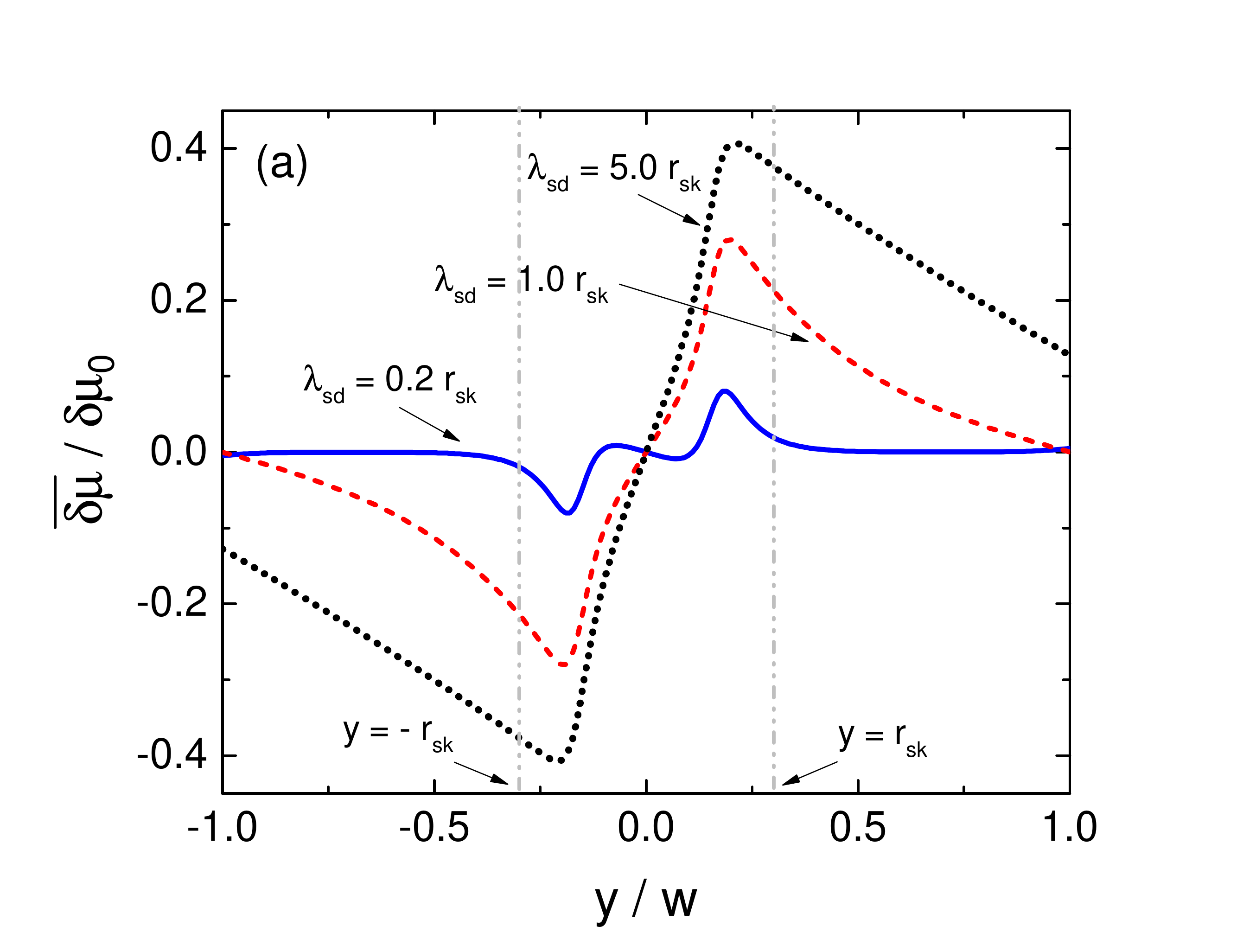}}
\hspace*{\fill}
\subfigure{
\includegraphics[trim={0.8cm 0.5cm 3cm 2.2cm},clip=true, width=0.85\columnwidth]{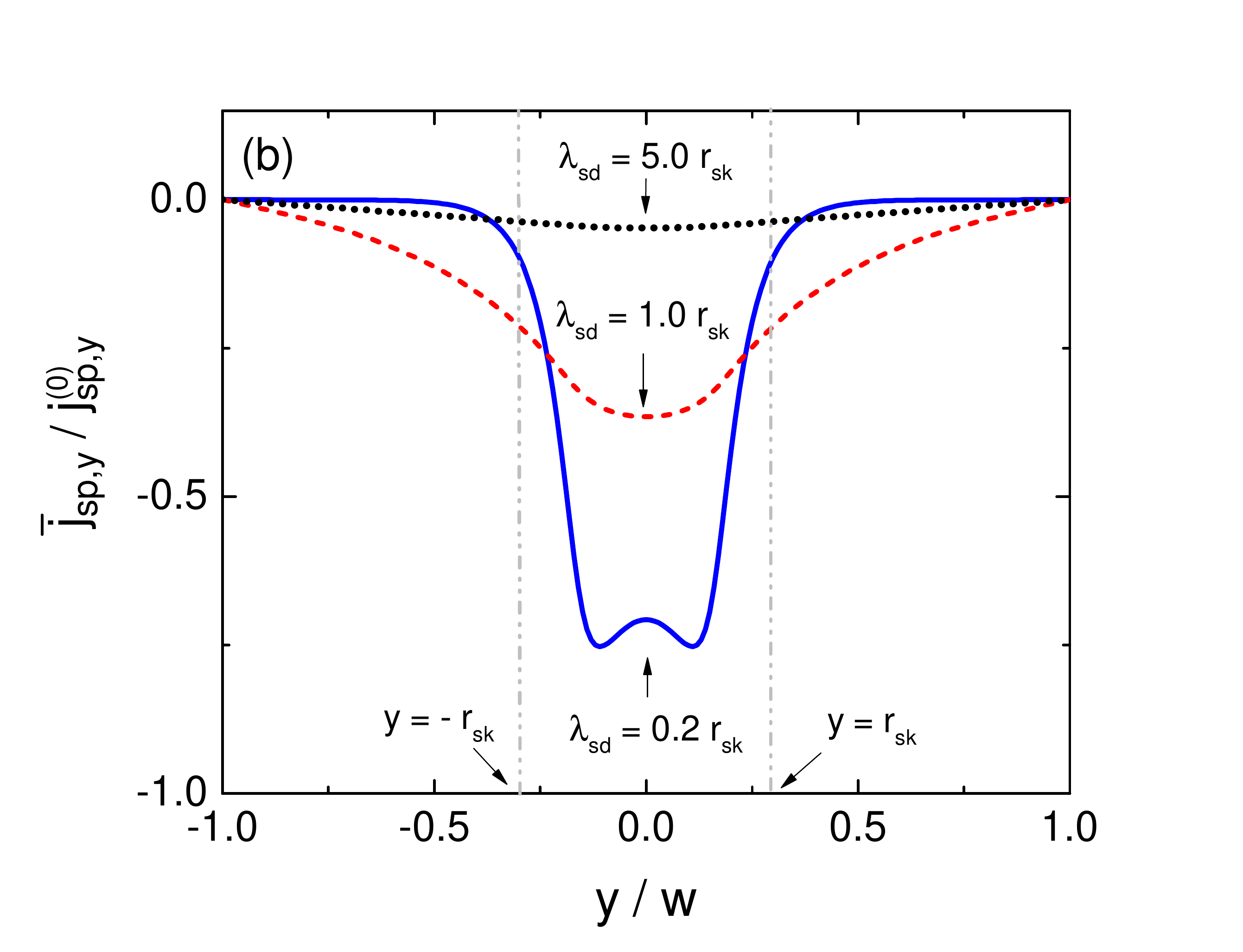}}
\hspace*{\fill}
\caption{Spatial profiles of (a) spin accumulation $\overline{\protect\delta
\protect\mu }$ and (b)transverse spin current density $j_{sp,y}$ in the
y-direction induced by a single skyrmion bubble for several different spin
diffusion lengths, where $\protect\delta \protect\mu _{0}\equiv E_{x}r_{sk}(%
\frac{e\protect\tau B_{0}}{m})$ and $j_{sp,y}^{(0)}\equiv (1-p_{\protect%
\sigma }^{2})\protect\sigma E_{x}(\frac{e\protect\tau B_{0}}{m})$, and the
skyrmion radius is taken to be $r_{sk}=\frac{w}{3}$. we have also used $%
a_{DW}=0.2r_{sk}$ and $J_{ex}=0.2\protect\epsilon _{F} $.}
\label{Fig:js-dmu_skyrm-bbl}
\end{figure}

In Fig.~\ref{Fig:js-dmu_skyrm-bbl}, we show the spatial distributions of the
spin accumulation $\overline{\delta \mu }$ and transverse spin current
density $\bar{j}_{sp,y}$ along the $y$-direction of the thin film (averaged
over the $x$ coordinate across the skyrmion). When the spin diffusion length
$\lambda _{sd}$ is much smaller than radius of the skyrmion bubble $r_{sk}$
(see the solid blue line for $\lambda _{sd}=a_{DW}=0.2~r_{sk}$), the
magnitudes of both $\overline{\delta \mu }$ and $\bar{j}_{sp,y}$ reach their
maxima around the narrow N\'{e}el wall, at variance with the case for a
single N\'{e}el skyrmion as shown in Fig.~\ref{Fig:js-dmu}. When $\lambda
_{sd}$ is comparable or larger than $r_{sk}$, the spatial profiles of $%
\overline{\delta \mu }$ and $\bar{j}_{sp,y}$ for a skyrmion bubble coincide
with those for a standard N\'{e}el skyrmion, since the characteristic range
of nonlocality, set by the spin diffusion length, spans the entire skyrmion
which makes the transport properties insensitive to the local spin structure
of the skyrmion.

\begin{figure}[h]
\includegraphics[trim={0.8cm 0.5cm 3cm 2.2cm},clip=true,
width=0.85\columnwidth]{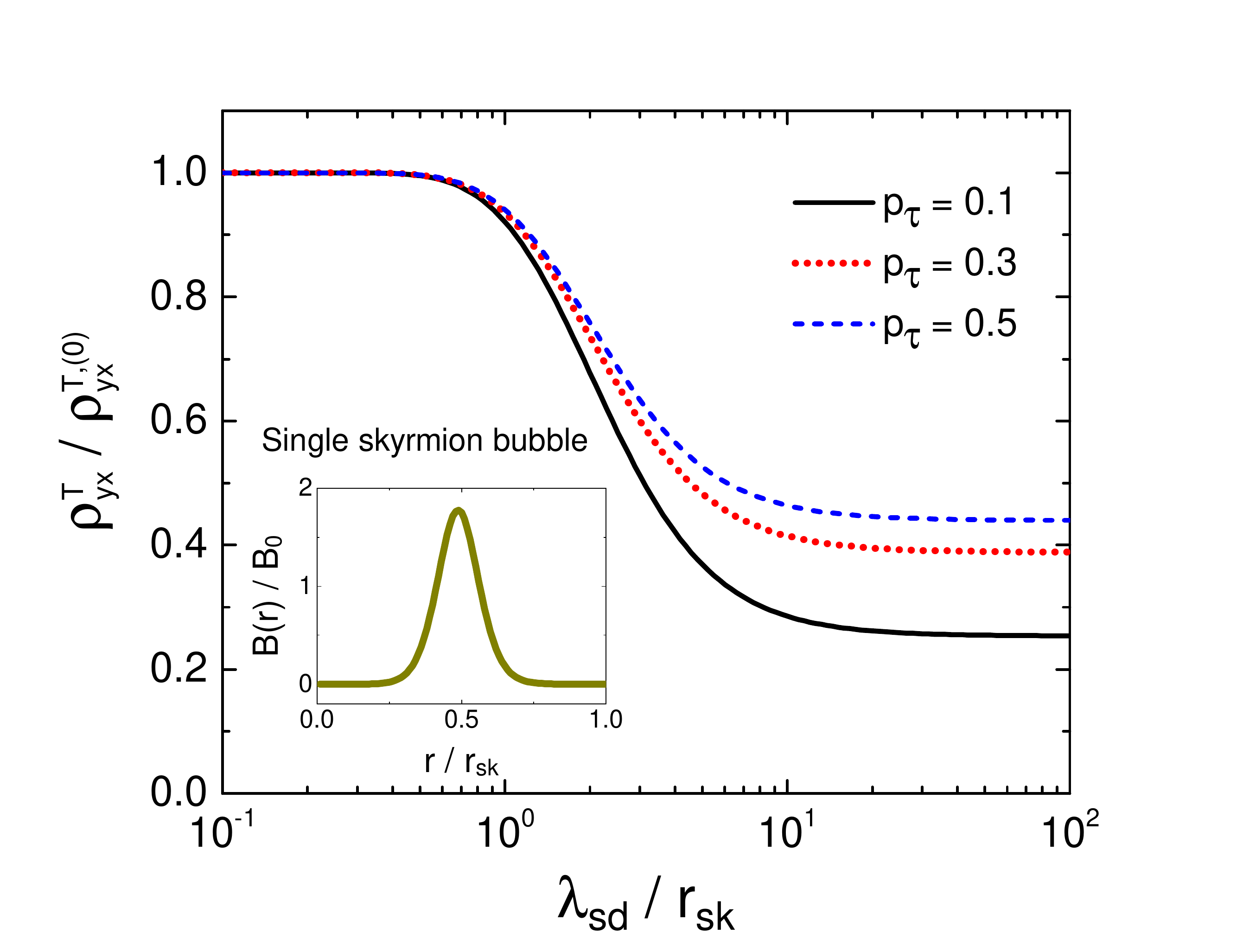} \hspace*{\fill%
}
\caption{Topological Hall resistivity $\protect\rho _{yx}^T$ generated by a
single skyrmion bubble as a function of spin diffusion length $\protect%
\lambda _{sd}$ in a thin film of width $w=6r_{sk}$ for several different $p_{%
\protect\tau }$, where $\protect\rho _{yx}^{T,(0)}$ is the ideal bulk TH
resistivity independent of $\protect\lambda _{sd}$; the insets show the
corresponding spatial profile of the emergent magnetic field. We have
defined $\protect\rho _{TH}^{\left( 0\right) }=\left( p_{\protect\tau }+p_{%
\protect\sigma }\right) R_{H}\protect\psi _{0}/S$ and used $a_{DW}=0.2r_{sk}$
and $J_{ex}=0.2\protect\epsilon _{F} $.}
\label{Fig:rho-1Skrm-bbl}
\end{figure}

In Fig.~\ref{Fig:rho-1Skrm-bbl}, we show the topological Hall resistivity as
a function of the ratio of $\lambda _{sd}/r_{sk}$. We find that the
topological Hall resistivity induced by a single skyrmion bubble turns out
to be exactly the same as that for a standard N\'{e}el skyrmion. This is
understandable since the topological Hall resistivity is calculated by
integrated over the entire skyrmion and hence only relies on the topology
(or topological charge) of the skyrmion rather than its detailed spin
structure.

\bibliographystyle{apsrev4-1}
\bibliography{180118_TH-spin-diffusion}

\begin{thebibliography}{40}
\bibitem [{pta()}]{ptau}%
  \BibitemOpen
  \href@noop {} {}\bibinfo {note} {Note that $p_{\sigma}=p_{\tau}+p_{n}$, where
  $p_n \simeq J_{ex}/\varepsilon_F$ with exchange splitting
  $J_{ex}=(\varepsilon_{F\uparrow}-\varepsilon_{F\downarrow})/2$ and the spin
  averaged Fermi energy
  $\varepsilon_{F}=(\varepsilon_{F\uparrow}+\varepsilon_{F\downarrow})/2$.}\BibitemShut
  {Stop}%
\end{thebibliography}

\end{document}